\documentstyle[prl,aps,epsf,prbbib]{revtex}
\newcommand{\gapprox}{\gtrsim}
\newcommand{\lapprox}{\lesssim}

\begin{document}
\draft
\twocolumn[\hsize\textwidth\columnwidth\hsize\csname  
@twocolumnfalse\endcsname

\title{{Magnetic Field Effect on the Pseudogap Temperature within
Precursor Superconductivity}}
\author{P. Pieri, G.C. Strinati, and D. Moroni}
\address{
Dipartimento di Matematica e Fisica, Sezione INFM
Universit\`{a} di Camerino, I-62032 Camerino, Italy}
\date{\today}
\maketitle
\hspace*{-0.25ex}

\begin{abstract}
We determine the magnetic field dependence of the pseudogap closing
temperature $T^{*}$ within a
precursor superconductivity scenario.
Detailed calculations with an anisotropic attractive Hubbard model account
for a recently
determined experimental relation in BSCCO between the pseudogap closing
field and the pseudogap
temperature at zero field, as well as for the weak initial dependence of
$T^{*}$ at low fields.
Our results indicate that the available experimental data are fully
compatible with a
superconducting origin of the pseudogap in cuprate superconductors.
\end{abstract}
\hspace*{-0.25ex}
]
\narrowtext

High-temperature superconductors are characterized by a {\em pseudogap
phase}, whereby a temperature
$T^{*}$ for the closing  of the pseudogap is identified in addition to the
(lower) superconducting critical
temperature $T_{c}$.
The origin of this pseudogap phase remains controversial and has mainly been
ascribed to two alternative scenarios, namely,
to the occurrence of a ``competing order parameter'' (due, e.g., to an
underlying antiferromagnetic or  charge
order)  or to the  presence of ``precursor superconductivity'' (due to
strong pairing attraction above $T_{c}$).
The issue is thus to decide whether pseudogap and superconductivity
(better, precursor pairing) represent
related or different phenomena.

In this context, magnetic effects are good candidates for distinguishing
between the two scenarios.
It turns out that $T^{*}$ and $T_{c}$ behave differently as functions of
hole concentration and
magnetic field $H$, in the sense that, in the underdoped region, $T^{*}$ is
weakly dependent on $H$
while $T_{c}$ shows more pronounced dependence \cite{Zheng-1999}.
A similar behavior is observed at optimal doping \cite{Gorny-1999}, while
in the overdoped region
the pseudogap behavior is strongly field dependent \cite{Zheng-2000}.
The above distinction between the two temperature (or energy) scales has
sometimes been quoted as
evidence that the two scales have different physical origin
\cite{Krasnov-2000,Krasnov-2001}.

Recently, by applying magnetic fields up to $60$ T, a systematic
determination of the pseudogap
closing field $H_{{\rm pg}}$ (above which the pseudogap phase is destroyed at 
any temperature) has been
made using data for
several doping values \cite{Shibauchi-2001}, showing strikingly different
dependencies from the
characteristic fields  of the superconducting state.
In addition, it has been found \cite{Shibauchi-2001} that $H_{{\rm pg}}$ and
$T^{*}$ at zero field are
related (within experimental uncertainty) through the simple Zeeman-like
expression
$g \mu_{B} H_{{\rm pg}} = k_{B} T^{*}$, where $g = 2$ is the electronic
$g$-factor, $\mu_{B}$ the Bohr
magneton, and $k_{B}$ the Boltzmann constant.
This expression suggests that the magnetic field couples to the pseudogap
by the Zeeman energy
of the spin degrees of freedom, thus entailing a predominant role of the
spins over the orbital
frustration effects in the formation of the pseudogap.

Within the framework of the pairing scenario, a calculation of the effect
of a magnetic field on
$T^{*}$ and $T_{c}$ has already been presented by including {\em orbital
\/} magnetic effects only
but no Zeeman coupling in a continuum model \cite{Levin-2001}, with the
conclusion that orbital
effects should contribute in an important way to the field dependence of
$T^{*}$.

In this Letter, we calculate the magnetic field dependence of $T^{*}$
within a precursor
superconductivity scenario, by taking explicit {\em account of the Zeeman
splitting for the spin
degrees of freedom}, and provide specific comparison with the data of Ref.~\onlinecite{Shibauchi-2001}.
To this end, we initially consider the Zeeman splitting only, within an
attractive Hubbard model for
an anisotropic (layered) structure with a $d$-wave pseudogap.
At a second stage, we include also orbital effects in a continuum model, to
assess the relative
importance of Zeeman and orbital couplings.
Our results are that, when pseudogap effects are weak (i.e., at low pairing
interaction), $T^{*}$
depends markedly on $H$, while at stronger pairing $T^{*}$ is weakly field
dependent.
In addition, at the doping levels relevant to experiments, we reproduce
quantitatively the expression $g \mu_{B} H_{{\rm pg}} = k_{B} T^{*}$ proposed
 by the authors of
Ref.~\onlinecite{Shibauchi-2001}.
Specifically, we find that: (i) Inclusion of the sole Zeeman
splitting for an
attractive Hubbard model with an anisotropic (layered) structure recovers
the above expression
within about a factor of two (which we attribute to the difference between
the temperature $T^{*}$ as
identified within precursor superconductivity and the energy scale
associated with pair breaking);
(ii) The contribution of the orbital effect (estimated for a
continuum model) does not significantly affect the results obtained for 
$H_{pg}$ with the sole inclusion of  the Zeeman effect, in the
intermediate-coupling regime relevant to experiments; (iii) In the small $H$ 
region, $T^*(H)$ is weakly dependent on $H$ and somewhat sensitive to the 
inclusion of the orbital effect.
Our results imply that the magnetic field effects in the pseudogap phase,
as determined by the experiments reported in Ref.~\onlinecite{Shibauchi-2001} 
as well as in Refs.\onlinecite{Zheng-1999,Gorny-1999}, and\onlinecite{LAO},
are fully compatible with precursor superconductivity as the origin of $T^{*}$.

The relative importance of the Zeeman and orbital effects on $H_{{\rm pg}}$ could
be assessed by applying
a magnetic field, alternatively, {\em parallel \/} and {\em orthogonal
\/} to the layered  structure
of the high-temperature superconducting materials, thus suppressing the
orbital contribution when the
magnetic field is parallel to the layers.
Although no experimental indication in this sense yet exists, the relation
between $H_{{\rm pg}}$ and
$T^{*}$ found in Ref.~\onlinecite{Shibauchi-2001} constitutes {\em per se\/} an
empirical suggestion
that the Zeeman effect dominates over the orbital frustration, at least as
far as $H_{{\rm pg}}$ is
concerned.

For the above reasons, we consider initially the Zeeman effect only, which
can be evaluated for
arbitrary magnetic field and for which lattice effects (even with
anysotropy) and the $d$-wave
character of precursor pairing can be readily included.
To this end, we consider an {\em attractive extended Hubbard model\/},
with nearest-neighbor
($t_{\parallel}$) and next-nearest neighbor ($t'_{\parallel}$) in-plane
hoppings, nearest-neighbor
($t_{\perp}$) out-of-plane hopping, and an in-plane attraction (with
strength $V \le 0$) between
nearest-neighbor sites.
This model is believed to capture the essential physics of cuprate
superconductors in the
nearly optimally doped region.

In the absence of magnetic field, a {\em pair breaking temperature\/} is
obtained by supplementing
the standard BCS gap equation with the particle-density equation to
determine the chemical
potential \cite{NSR}, since its value is no longer pinned at the Fermi
level as soon as the
attraction is sufficiently strong.
When this happens, the above temperature is no longer related to the
superconducting critical
temperature, but rather signals the formation of (quasi) bound pairs.
In this context, the pair breaking temperature is then associated with the
crossover temperature
$T^{*}$ identified experimentally.
In the following, we extend this identification to the presence of a
magnetic field.
The coupled equations to be solved are:

\begin{equation}
\frac{1}{|V|} \, = \, \int \! \frac{d{\bf  k}}{(2\pi)^{3}} \,\,\,
\frac{\gamma({\bf  k})^{2}}{4 \, \xi({\bf  k})} \,\, \sum_{\sigma} \,\,
\tanh \left(\frac{\xi_{\sigma}({\bf  k})}{2\, k_{B} \, T^{*}} \right)
\label{gap-equation}
\end{equation}

\begin{equation}
n \, = \, \, \int \! \frac{d{\bf  k}}{(2\pi)^{3}} \,\, \sum_{\sigma} \,\,
\frac{1}{\exp (\xi_{\sigma}({\bf  k})/k_{B}\, T^{*}) \, + \, 1} \,\,\, .
\label{density-equation}
\end{equation}

\noindent
In these expressions, $n$ is the site density, $\xi_{\sigma}({\bf  k}) =
- 2t_{\parallel} (\cos k_{x} + \cos k_{y}) - 2 t_{\perp} \cos k_{z}
+ 4 t'_{\parallel} \cos k_{x} \cos k_{y} - \mu + \mu_{B}H\sigma$
(where $\mu$ is the chemical potential and $\sigma = \pm 1$),
$2 \xi({\bf  k}) = \xi_{+}({\bf  k}) + \xi_{-}({\bf  k})$,
$\gamma({\bf  k}) = (\cos k_{x} - \cos k_{y})$ for a $d$-wave gap, and each
component of the wave vector ${\bf  k}$ is integrated from $-\pi$ to $+\pi$.
To make contact with experimental data, we interpret $n = 1 - p$ where $p$
is the
doping value \cite{Shibauchi-2001}.
Values appropriate to BSCCO are $t_{\parallel} = 0.25$ eV, $t'_{\parallel} =
0.45 t_{\parallel}$,
and $t_{\perp} = 0.2t_{\parallel}$ \cite{Pavarini}.
The value of $|V|$ (corresponding to a given sample) is obtained by solving
the above equations
for $H=0$ and setting $T^{*}(H=0)$ at the experimental value in zero field.
These equations are then solved numerically for the unknowns $(T^{*},\mu)$
with given values of
$(n,|V|,H)$.
In this way, to any given sample (corresponding to a pair of values $(n,|V|)$),
we associate a curve $T^{*}(H)$, from which the pseudogap closing field
$H_{{\rm pg}}$ of interest
is extracted by extrapolation to $T^{*}(H_{{\rm pg}})=0$ (extrapolation is
required to avoid numerical
instabilities when $T^{*}(H)$ approaches zero).

A typical curve $T^{*}(H)$ obtained by the above procedure is shown in Fig.~1,
for the values of $n$ and $|V|$ corresponding to the sample with $p=0.21$
analyzed in
Ref.~\onlinecite{Shibauchi-2001}.
Note the {\em rather weak initial dependence\/} of $T^{*}(H)$ for low $H$.
Typically, $T^{*}(H)$
decreases only by about $1\%$ for values of $H$ up to $25$ T.
This result is consistent with the experimental finding reported by several
authors
\cite{Zheng-1999,Gorny-1999,LAO}, where a negligibly weak magnetic-field
dependence of $T^{*}$
is noted for various compounds.

\begin{figure}
\centering
\narrowtext 
\epsfxsize=3.1in
\epsfbox{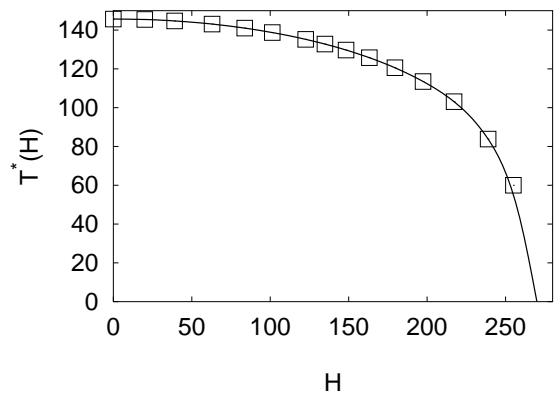}
\vspace{.1truecm}
\caption{
Dependence of $T^{*}(H)$ on $H$ obtained from
Eqs.~(\ref{gap-equation}) and
(\ref{density-equation}) with $n=0.79$ and $|V|=0.175$ eV.
The continuous curve is a polynomial fit.
[Temperature is expressed in Kelvin and magnetic field in Tesla.]
}
\end{figure}

Figure~2 shows the results of our calculation corresponding to the three
samples reported in the
inset of Fig.~4 from Ref.~\onlinecite{Shibauchi-2001}.
In that reference, the large ($\gapprox 100$ T) values of $H_{{\rm pg}}$  have
been obtained by extrapolation
of the data collected up to $60$ T.
Note that both the experimental and the theoretical points lie on straight
lines passing through
the origin.
This entails a {\em linear proportionality\/} between $H_{{\rm pg}}$ and
$T^{*}(H=0)$ (of the form
$2 \mu_{B} H_{{\rm pg}} = \alpha k_{B} T^{*}(H=0)$ where $\alpha$ is a constant),
as pointed out in
Ref.~\onlinecite{Shibauchi-2001}.
[This linear proportionality has ben established in
Ref.~\onlinecite{Shibauchi-2001} not only for the
three doping values reported in the inset of their Fig.~4, but also for
lower doping values (albeit
with larger uncertainty), as shown in their Fig.~4.
We have also extended our calculation to the value $p=0.16$, still obtaining 
the linear behavior.]

\begin{figure}
\centering
\narrowtext 
\epsfxsize=3.1in
\epsfbox{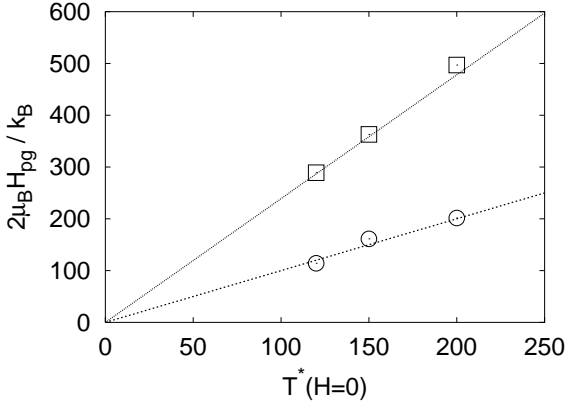}
\vspace{.1truecm}
\caption{
Theoretical (squares) and experimental (circles) values of
$2\mu_{B}H_{{\rm pg}}/k_B$ vs $T^{*}(H=0)$, for the three samples [with 
$p=(0.20,0.21,0.22)$ from right to left] reported in the inset of Fig.~4 from 
Ref.~6. [Temperatures are expressed in Kelvin.]} 
\end{figure}

The experimental data give $\alpha_{{\rm{{\rm exp}}}} \simeq 1$, while our
calculation yields
$\alpha_{{\rm theo }} \simeq 2.4$.
This difference should not be regarded to be too severe.
First of all, recall that $T^{*}$ is, by its own nature, a non-sharply
defined crossover
temperature, so that there is an intrinsic uncertainty in its identification.
Nonetheless, one may attribute the above discrepancy to additional physical
effects (such as
orbital frustration) not included in Eqs.~(\ref{gap-equation}) and
(\ref{density-equation}).
We have, however, verified for the continuum model (see below) that orbital
frustration
contributes only in a marginal way to the suppression of $T^{*}$ in the
intermediate-coupling
regime.
In addition, one may observe that, while the experimental value
$T^{*}_{{\rm exp}}$ (as obtained
in Ref.~\onlinecite{Shibauchi-2001}) identifies directly an energy scale 
(which, by our interpretation,
is associated with the binding energy $E_{b}$ of a pair embedded in the
medium \cite{Kyung}),
the theoretical value $T^{*}_{{\rm theo}}$ can be {\em smaller\/} than $E_{b}$.
This statement is readily verified in the strong-coupling limit of the
continuum model considered
below, whereby $k_{B} T^{*}_{{\rm theo}} = (\epsilon_{o}/2) / \ln
(\epsilon_{o}/\epsilon_{F})^{3/2}$
in zero field \cite{PE}, $\epsilon_{F}$ being the Fermi energy and
$\epsilon_{o}$ the binding energy
of the associated two-body problem ($E_{b}$ is expected to reduce to
$\epsilon_{o}$ in this limit).
From this relationship, one gets $T^{*}_{{\rm theo}} < \epsilon_{o}$ when
$\epsilon_{o}/\epsilon_{F} > \exp (1/3) \simeq 1.4$.
At weaker (intermediate) couplings, on the other hand, numerical
calculation yields
$T^{*}_{{\rm theo}} \simeq \epsilon_{o}/2$ when $\epsilon_{o}/\epsilon_{F} \simeq 1$.
In this case, the experimental relation $g \mu_{B} H_{{\rm pg}} = k_{B} T^{*}_{{\rm exp}}$
is thus equivalent to the theoretical relation $g \mu_{B} H_{{\rm pg}} =
\alpha_{{\rm theo}} k_{B} T^{*}_{{\rm theo}}$
with $\alpha_{{\rm theo}} \simeq 2$ (provided one still identifies $E_{b}$ with
$\epsilon_{o}$).
The difference between the experimental and theoretical identification of
$T^{*}$ may thus
account for the difference between $\alpha_{{\rm exp}}$ and $\alpha_{{\rm theo}}$.

In any event, the physical point to be emphasized from our calculation is
that the
{\em energy scales\/} associated with $H_{{\rm pg}}$ and $T^{*}(H=0)$ are of the
{\em same order\/},
as found experimentally.
The experimental result thus finds a natural explanation within our
precursor superconductivity
scenario, whereby $T^{*}(H=0)$ is associated with the binding energy of the
(quasi) preformed pairs
and $H_{{\rm pg}}$ with the Pauli pair breaking due to Zeeman splitting.

To assess the role of orbital frustration, it is sufficient to analyze the
simpler continuum model
for which both Zeeman and orbital effects can be readily included.
In this way, we avoid facing at this stage the peculiar problems of Bloch
electrons in a magnetic
field \cite{Harper,Hofstadter,MS}.

In the continuum case, the fermionic attraction can be modeled by a point
contact interaction,
regularized (in three dimensions) in terms of the scattering length $a_{F}$
\cite{Randeria-1993,Pi-S-98}.
The dimensionless interaction parameter $(k_{F} a_{F})^{-1}$ (where the
Fermi wave vector $k_{F}$
is related to the density via $n=k_{F}^{3}/(3\pi^{2})$) then ranges from
$-\infty$ in weak coupling
to $+\infty$ in strong coupling (the crossover region of interest being
bound by
$(k_{F} |a_{F}|)^{-1} \lapprox 1$).
For this model, the equations identifying $T^{*}$ within our approach (for
an $s$-wave pseudogap) and
corresponding to Eqs.~(\ref{gap-equation}) and (\ref{density-equation}) are
the following:

\begin{eqnarray}
& &\frac{m}{4 \pi a_{F}} +  R(T^{*},\mu;H) - \omega_{c}\, 
b(T^{*},\mu;H) = 0
\label{gap-equation-continuum}\\
& & n = \frac{m\omega_{c}}{2 \pi} \sum_{s=0}^{\infty} 
\int_{-\infty}^{+\infty} \! \frac{dk_{z}}{2\pi} \sum_{\sigma} 
f\left[\left(s+\frac{1}{2}\right)\omega_{c} + \frac{k_{z}^{2}}{2m} -
\mu_{\sigma}\right]
\label{density-equation-continuum}
\end{eqnarray}

\noindent
where $m$ is the fermion mass, $\omega_{c} = e H/(m c)$ the cyclotron
frequency,
$\mu_{\sigma} = \mu - \mu_{B} H \sigma$, $f(x) = [\exp (x/k_{B} \,
T^{*}) + 1 ]^{-1}$ the
Fermi function at temperature $T^{*}$, and

\begin{eqnarray}
R(T^{*},\mu;H) &=& \int_{0}^{\infty} \! d\epsilon \, N(\epsilon)
\left[\sum_{\sigma}\frac{
\tanh \left(\frac{\epsilon - \mu_{\sigma}}{2  k_{B} T^{*}} \right)}
{4(\epsilon - \mu)}
 - \frac{1}{2\epsilon}  \right]
\label{R}\\
b(T^{*},\mu;H) &=& \int_{0}^{\infty} \! d\epsilon \,
\frac{N(\epsilon)\, \epsilon}{6(\mu-\epsilon)^2}\left[\frac{F(\epsilon)}{(\mu-\epsilon)}+
F'(\epsilon)\right]\,\, ,
\label{b}
\end{eqnarray}

\noindent
$N(\epsilon)$ being the independent-particle density of states per spin
component and
$F(\epsilon) = f(\epsilon - \mu_{+}) + f(\epsilon - \mu_{-}) -1$.
These equations contain both the Zeeman splitting (through the spin
dependence of $\mu_{\sigma}$)
{\em and\/} the orbital effect (through the presence of $b(T^{*},\mu;H)$
in Eq.~(\ref{gap-equation-continuum})
and of the Landau levels in Eq.~(\ref{density-equation-continuum})).
These equations include, however, the effect of the magnetic field on a
different footing.
Equation (\ref{gap-equation-continuum}) (which has been obtained by
considering the Thouless
instability in the particle-particle channel) bears on the eikonal
approximation, whereby the
orbital effect of the magnetic field is introduced by multiplying the
real-space representation
of the independent-particle Green's function in the absence of the field by
a phase factor
$\exp \{i \varphi \}$ \cite{Gorkov}.
Equation (\ref{density-equation-continuum}), on the other hand, keeps the
Landau orbital quantization
explicitly, otherwise the eikonal approximation would completely wash out
the  magnetic orbital
effect.

It is relevant to discuss at this point the conditions for the  validity of
the eikonal approximation
in the present context.
The domain of validity of the eikonal approximation for the single-particle
Green's function
depends on the distance between its two spatial coordinates.
For Eq.~(\ref{gap-equation-continuum}), this domain is delimited by the
range $l$ of the two-particle
propagator in zero field, yielding $\omega_{c} \ll p_{G}/(m l)$ where
$p_{G}$ is largest momentum scale
of the independent-particle Green's function.
In addition, by expanding up to second order the phase factor $\exp \{i
\varphi \}$ of the eikonal
approximation in Eq.~(\ref{gap-equation-continuum}), one arrives at the
additional condition
$\omega_{c} \ll (m l^{2})^{-1}$.
In weak coupling, $p_{G} \approx k_{F}$ and $l \approx \xi_{o}$ where
$\xi_{o}$ is the Pippard
coherence length.
In this case, one obtains $\omega_{c} \ll (m \xi_{o}^{2})^{-1}$ for the
validity of the eikonal
approximation.
In strong coupling, on the other hand, $p_{G}^{-1} \approx a_{F} \approx
l$, leading to
$\omega_{c} \ll \epsilon_{o}$ for the validity of the eikonal approximation.

The above estimates lead us to conclude that, strictly speaking, the
eikonal approximation holds
only for the {\em initial part\/} of the curve $T^{*}(H)$ close to zero field.
Taking $\omega_{c} m \xi_{o}^{2} \lapprox 0.1$ in weak and
$\omega_{c}/\epsilon_{o} \lapprox  0.1$
in strong coupling, one obtains correspondingly
$(T^{*}(H=0) - T^{*}(H))/T^{*}(H=0)$ $\lapprox 1 \%$ and $\lapprox 10 \%$.
The relative range of validity of the eikonal approximation increases,
therefore, when passing from
weak to strong coupling.
In practice, we shall extrapolate the results for $T^{*}(H)$ obtained from
Eqs.~(\ref{gap-equation-continuum}) and (\ref{density-equation-continuum})
outside the strict
range of validity of the eikonal approximation to obtain the pseudogap
closing field $H_{{\rm pg}}$,
since we are interested in obtaining a qualitative estimate of the relative
importance of
the Zeeman and orbital effects at intermediate coupling.
The eikonal approximation, on the other hand, does not affect the equations
determining
$T^{*}(H)$ when the Zeeman effect only is included.

\begin{figure}
\centering
\narrowtext 
\epsfxsize=3.1in
\epsfbox{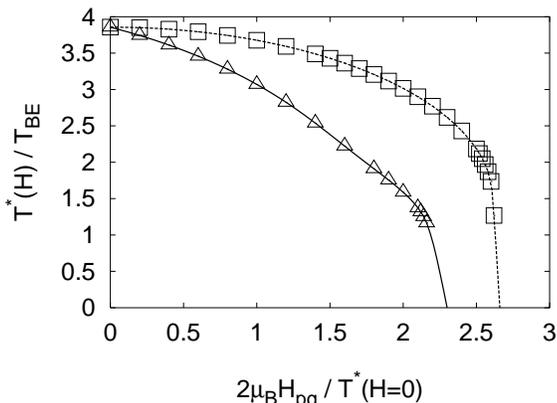}
\vspace{.1truecm}
\caption{
$T^{*}(H)$ vs $H$ for the continuum model with coupling value
$(k_{F}a_{F})^{-1}=0.5$, including the Zeeman effect only (squares) and
the orbital effect as well (triangles). The continuous curves are polynomial
fits.
}
\end{figure}

Figure~3 shows $T^{*}(H)$ vs $H$ obtained by
Eqs.~(\ref{gap-equation-continuum}) and
(\ref{density-equation-continuum}) for the continuum model with the value
$(k_{F}a_{F})^{-1}=0.5$
in the intermediate-coupling regime, by including both Zeeman {\em and\/}
orbital effects (triangles)
or the Zeeman effect only (squares) (the temperature $T^{*}$ has been
conveniently normalized in
terms of the Bose-Einstein temperature $T_{BE}$ for composite bosons of
mass $2m$ and density
$n/2$).
Note that $\alpha_{{\rm theo}} \simeq 2.6$ with the Zeeman effect only, while
$\alpha_{{\rm theo}} \simeq 2.3$
when both effects are included (to be compared with the value
$\alpha_{{\rm theo }} \simeq 2.4$
from Fig.~2).
This implies that the calculation with the Zeeman effect only is rather
stable against the inclusion
of the orbital effect.
Note also from this figure that, in the small $H$ region, $T^{*}(H)$ is
somewhat sensitive to
the inclusion of the orbital effect (albeit the dependence of $T^*(H)$ on $H$ 
is still weak in an absolute sense), in agreement with the experimental
result\cite{LAO} for fields up to $14$ T.

In conclusion, we have shown that consideration of the Zeeman splitting
within the precursor
superconductivity scenario (at intermediate coupling) reproduces the
experimental relation
$2 \mu_{B} H_{{\rm pg}} = k_{B} T^{*}(H=0)$ between the pseudogap closing field
$H_{{\rm pg}}$ and the pseudogap
temperature $T^{*}(H=0)$ at zero field \cite{Shibauchi-2001}.
Our results are also fully consistent with the occurrence of a weak initial
dependence of $T^{*}(H)$
on $H$ at low fields \cite{Zheng-1999,Gorny-1999,LAO}.
As the sensitivity to a magnetic field is believed to be a crucial test for
the physical origin of
the pseudogap, our results show that the available experimental data are
fully compatible with a
superconducting origin of the pseudogap.
\acknowledgments

We are indebted to D. Neilson and A. Perali for useful discussions.

%%%%%%%%%%%%%%%%%%%%%%%%%%%%%%%%%%%%%%%%%%%%%%%%%%%%%%%%%%%%%%%%%%%%%%%%%%%%
%%%%%%%%%%%%%%%%%%%%%%%%%

% Bibliography

%%%%%%%%%%%%%%%%%%%%%%%%%%%%%%%%%%%%%%%%%%%%%%%%%%%%%%%%%%%%%%%%%%%%%%%%%%%%
%%%%%%%%%%%%%%%%%%%%

\end{document}